# A Comparison of Strategies to Embed Physics-Informed Neural Networks in Nonlinear Model Predictive Control Formulations Solved via Direct Transcription

Carlos Andrés Elorza Casas, Luis A. Ricardez-Sandoval, and Joshua L. Pulsipher[1]

*Department of Chemical Engineering, University of Waterloo, Waterloo, ON N2L 3G1, Canada*

**Abstract**

This study aims to benchmark candidate strategies for embedding neural network (NN) surrogates in nonlinear model predictive control (NMPC) formulations that are subject to systems described with partial differential equations and that are solved via direct transcription (i.e., simultaneous methods). This study focuses on the use of physics-informed NNs and physics-informed convolutional NNs as the internal (surrogate) models within the NMPC formulation. One strategy embeds NN models as explicit algebraic constraints, leveraging the automatic differentiation (AD) of an algebraic modelling language (AML) to evaluate the derivatives. Alternatively, the solver can be provided with derivatives computed external to the AML via the AD routines of the machine learning environment the NN is trained in. The three numerical experiments considered in this work reveal that replacing mechanistic models with NN surrogates may not always offer computational advantages when smooth activation functions are used in conjunction with a local nonlinear solver (e.g., Ipopt), even with highly nonlinear systems. Moreover, in this context, the external function evaluation of the NN surrogates often outperforms the embedding strategies that rely on explicit algebraic constraints, likely due to the difficulty in initializing the auxiliary variables and constraints introduced by explicit algebraic reformulations.

---

[1] Corresponding author information: Phone: 1-519-888-4567 ext. 42290. Fax: 1-519-888-4347
E-mail: pulsipher@uwaterloo.ca

## 1. Introduction

Nonlinear model predictive control (NMPC) has become a widely accepted control algorithm to deal with constrained multivariable problems [1], [2], [3], [4]. Large-scale chemical processes pose an important challenge to the implementation of NMPC. Models for industrial-scale applications are often large, highly nonlinear, and involve multiple inputs, outputs, and states [5]. Systems described by nonlinear partial differential equations (PDEs) are especially challenging since they often produce formulations with multiple states. In this context, PDEs (which are infinite-dimensional modelling objects [6]) are often converted into algebraic constraints that are compatible with conventional nonlinear programming (NLP) solvers [7], [8]. This conversion is accomplished via direct transcription, which discretizes the PDEs by applying finite difference or collocation methods [9]. This approach results in a large system of nonlinear algebraic equality constraints which the NLP solver can solve simultaneously with the optimization of the NMPC problem.

Most gradient-based NLP solvers (e.g., Ipopt [1]) require the gradients of the objective and constraints with respect to the decision variables [10]. Although direct transcription approaches often generate optimization formulations with a large number of variables and constraints, the gradients can efficiently be evaluated through the automatic differentiation (AD) provided by algebraic modelling languages (AMLs), such as JuMP [11] and Pyomo [12]. AMLs also leverage the specialized algebraic forms typically exhibited by NLPs to efficiently compute sparse second-order derivatives that accelerate the convergence of NLP solvers [10], [13], [14]. Despite these advantages, the computational cost of solving transcribed PDE-constrained NMPC problems can become prohibitive for online feedback control which requires the NMPC problem to be solved at every sampling interval. Model simplifications, such as linearization, can reduce the complexity of the problem, but such models may not be sufficiently accurate to achieve effective control, particularly for highly nonlinear systems.

In an effort to reduce the computational burden, machine learning (ML) surrogate models based on neural networks (NNs) are becoming a particularly popular choice to replace computationally intensive mechanistic models since NNs have universal approximation properties and are often fast to evaluate [15], [16], [17], [18]. Moreover, data-driven models such as NNs are commonly used in hybrid modelling techniques to achieve a higher accuracy than could be achieved with traditional mechanistic models alone [19]. Physics-informed NNs (PINNs) are commonly used as ML surrogates since they use fundamental equations to decrease (or remove) the need for historical/process data for training and increase the fidelity of the NN approximator relative to physical laws [15], [20]. Several control-oriented PINN architectures have been proposed. A few approaches have used PINNs to directly solve the optimal control problem by including the objective function within the loss function [21], [22]. However, the solution encoded in the trained PINN is only optimal for a given initial condition and must be retrained when the initial condition is updated, making such approaches prohibitively expensive for NMPC. Other studies have the PINN model act as surrogate state-space model that maps current states and control actions to states at the next sampling interval which avoids the need for retraining; such studies have used a variety of NN architectures such as recurrent NNs (RNNs), long-short term memory (LSTM) networks, and convolutional NNs (CNNs) [15], [23], [24], [25], [26], [27]. Some of the challenges that arise is how to structure the data and the ML model to accurately capture the state transitions. For example, a few works have kept time as a continuous input to a NN [15], [28], [29]. However, this requires training over more data points since the data must capture the time-domain, the state-space and the control-input space. Other studies have fully-discretized the differential equations modelling the system. Hence, the data must only capture the state-space and control-input space [24], [25], [26], [27]. Nevertheless, the accuracy of the model is limited by the accuracy of the discretization scheme.

When embedding a PINN into an NMPC problem that is solved via shooting with an ordinary differential equation (ODE) integrator, Antonelo et al. proposed a strategy that showed a small computational improvement relative to using the analytic/mechanistic model on small ODE-constrained problems [15]. That same approach was later applied to different ODE-constrained case studies that were still relatively small [28], [29]. Other studies have proposed methods that embed NN surrogate models into NMPC problems, but they do not report the optimization approach and/or the solution times. To name a few, Chen et al. used an RNN to learn the system dynamics and embedded the trained RNN into NMPC by evaluating the gradients via backpropagation [23]. Zheng et al. used physics-informed RNN surrogates to capture nonlinear systems subject to Gaussian measurement noise, and they provided the derivative information to Ipopt via finite difference approximations; however, they do not report the computational times [27]. Alhajeri et al. used LSTM surrogates for NMPC, but they do not report the embedding strategy or the computational times [24]. Hence, there is gap in the literature on determining what strategies are most effective for embedding NNs in NMPC formulations, especially when direct transcription is considered as the solution strategy to solve the NMPC problem.

One approach that has become popular with tools such as the Optimization and Machine Learning Toolkit (OMLT), is to embed NN surrogates via explicit algebraic constraints, which we refer to as Explicit Constraint Embedding (ECE) [30]. These constraints, along with the rest of the NMPC problem, are given to the AML, which provides gradients to the NLP solver via tailored AD routines. Alternatively, NN surrogates can be treated as external functions in the AML where the gradients are provided using the AD employed by the underlying ML environment used to train the NN (e.g., PyTorch [31] and Tensorflow [32]). In this approach, which we refer to as External Function Embedding (EFE), the NLP solver passes the current iteration of the decision variables to the ML environment which then applies the chain rule over the layers of the NN to evaluate the gradients of the external function that wraps the NN in the AML. To the authors' knowledge, these two approaches have not been juxtaposed in the literature for NMPC problems. Similarly, there are no computational studies available that use either approach on NMPC problems solved via direct transcription.

This work investigates the performance of ECE and EFE as strategies for embedding NNs into PDE-constrained NMPC problems solved using the direct transcription dynamic optimization method. The main contributions of our work include:
- extending the methodology proposed in [15] to leverage PINNs and physics-informed CNNs (PICNNs) to handle PDE-constrained NMPC problems, and
- benchmarking ECE against EFE for NMPC problems that employ NN models and use direct transcription.

The benchmarking work is based on three nonlinear plug-flow reactor (PFR) case studies of increasing complexity that encode typical PDE-constrained NMPC applications in chemical engineering. While the use of embedding strategies using shooting methods is outside the scope of this work, a simple shooting method based on the one used in [15] is included as a point of comparison.

This work is structured as follows: Section **Error! Reference source not found.** details the NMPC problems considered in this work. Section **Error! Reference source not found.** describes techniques used to develop and embed PINN/PICNN surrogates in the NMPC problems. Section **Error! Reference source not found.** details the benchmarking studies and discusses key results. Section 5 concludes with the core findings and suggests future work.

## 2. NMPC Problem Definition

This section introduces the nomenclature and formulations for the PDE-constrained NPMC problems that define the scope of this benchmarking study.

### 2.1. Infinite-Dimensional Formulation

We consider a PDE-constrained problem indexed by time $t \in \mathcal{T} = [t_0, t_f] \subset \mathbb{R}$ and spatial position $\mathbf{z} \in \Omega \subset \mathbb{R}^{n_z}$ with state variables $\mathbf{x}: \mathcal{T} \times \Omega \to \mathcal{X}$ ($\mathcal{X} \in [\mathbf{x}^L, \mathbf{x}^U] \subset \mathbb{R}^{n_x}$) and control variables $\mathbf{u}: \mathcal{T} \to \mathcal{U}$ ($\mathcal{U} \in [\mathbf{u}^L, \mathbf{u}^U] \subset \mathbb{R}^{n_u}$):

$$\min_{\mathbf{u},\mathbf{x}} \int_{\mathcal{T} \times \Omega} \|\mathbf{x}_{sp} - \mathbf{x}\|_L^2 + \|\mathbf{u}\|_W^2 \, d\mathbf{z}dt \tag{1a}$$

$$\text{s.t. } \frac{\partial \mathbf{x}}{\partial t}(t, \mathbf{z}) = \mathcal{F}_{\mathbf{z}}(\mathbf{x}, \mathbf{u})(t, \mathbf{z}), \quad t \in \mathcal{T}, \mathbf{z} \in \Omega \tag{1a}$$
$$\mathbf{g}(\mathbf{x}, \mathbf{u})(t, \mathbf{z}) \leq 0, \quad t \in \mathcal{T}, \mathbf{z} \in \Omega \tag{1a}$$
$$\mathbf{x}^L \leq \mathbf{x} \leq \mathbf{x}^U, \quad t \in \mathcal{T}, \mathbf{z} \in \Omega \tag{1a}$$
$$\mathbf{u}^L \leq \mathbf{u} \leq \mathbf{u}^U, \quad t \in \mathcal{T} \tag{1a}$$
$$\mathbf{u} = \mathbf{u}(t_M), \quad t \in [t_M, t_f] \tag{1a}$$
$$\mathcal{B}(\mathbf{x}, \mathbf{u})(t, \mathbf{z}) = 0, \quad t \in \mathcal{T}, \mathbf{z} \in \partial\Omega \tag{1a}$$
$$\mathcal{I}(\mathbf{x})(t_0, \mathbf{z}) = 0, \quad \mathbf{z} \in \Omega \tag{1a}$$

where $\mathbf{x}_{sp}: \mathcal{T} \times \Omega \to \mathcal{X}$ is the setpoint, $\mathbf{g}: \mathcal{X} \times \mathcal{U} \to \mathbb{R}^{n_g}$ are path constraint functions, $\partial\Omega$ is the boundary of $\Omega$, $\mathcal{F}_{\mathbf{z}}$ is a nonlinear spatial differential operator, $\mathcal{B}$ is the boundary condition operator, and $\mathcal{I}$ is the initial condition operator evaluated at the initial time $t_0$. The control actions $\mathbf{u}$ often appear as boundary conditions (e.g., at the inlet of a PFR) or as forcing terms within $\mathcal{F}_{\mathbf{z}}$. The PDE constraints in equation (1a) in general are complex nonlinear infinite-dimensional objects which are functions of multiple continuous independent variables in the time and spatial dimensions. The term $\|\mathbf{e}\|_B^2$ denotes the L$_2$-norm of vector $\mathbf{e}$ weighted by matrix $\mathbf{B}$ squared ($\|\mathbf{e}\|_B^2 = \mathbf{e}^T \mathbf{B} \mathbf{e}$). $\mathbf{L}$ and $\mathbf{W}$ are the output and input weight matrices, respectively, in the objective function. Often, the control actions can only change over some shorter time horizon, $t_M$ ($t_0 \leq t_M \leq t_f$), known as the control horizon, as shown in (1a). Problem (1a) cannot be solved directly by conventional solvers and must first be reformulated as described next.

## 2.2. Transcribed Formulation

To make Problem (1a) tractable with conventional solvers (e.g., Ipopt [14]), we transform it via direct transcription. That is, the time domain is discretized over the sampling interval $[0, \Delta t]$ to yield the set of time points $\hat{\mathcal{T}} = \{t_0 + k\Delta t : k \in \{0, 1, \ldots, P\}, t_f = P\Delta t\}$ where $P$ is the prediction horizon in the NMPC. The spatial domain is also discretised over a set of finite elements $\hat{\Omega} = \{z_v : v \in \{1, \ldots, n_{fe}\}\}$ where $n_{fe}$ is the number of finite elements (i.e., discretization points). Hence, $\hat{x}_{k,v} \in \mathbb{R}^{n_x}$ represents the state variables at time point $k$ and spatial location $v$, and it approximates the solution of the PDE at $x(t_k, z_v)$. Note that for compactness, the notation only considers a 1D discretized spatial domain $\hat{\Omega} \subset \mathbb{R}$ such that $\hat{x} \in \mathbb{R}^{n_x \times n_{fe}}$, but this work can be readily extended to higher dimensional domains (e.g., $\hat{x} \in R^{n_x \times n_{fe} \times n_{fe}}$). Similarly, $\hat{u}_k \in \mathbb{R}^{n_u}$ is the vector of control inputs at time $t_k$ which directly corresponds to $u(t_k)$ assuming that $u(t)$ is taken to be a piecewise constant function that is held constant over each sampling interval. By projecting Problem (1a) on $\hat{\mathcal{T}}$ and $\hat{\Omega}$, we obtain the discrete-time NMPC formulation:

$$\min_{\hat{u},\hat{x}} \sum_{k \in \{0,\ldots,P\}} \sum_{v \in \{1,\ldots N_{fe}\}} \eta_{k,v} \left( \|x_{sp}(t_k, z_v) - \hat{x}_{k,v}\|_2^2 + \|\hat{u}_k\|_2^2 \right) \tag{2a}$$

s.t.
$$\frac{\partial \hat{x}_{k+1,v}}{\partial t} = \hat{\mathcal{F}}_z(\hat{x}_{k,v}, \hat{u}_k), \quad k \in \{0, \ldots, P\}, v \in \{1, \ldots, n_{fe}\} \tag{2a}$$
$$g(\hat{x}_{k,v}, \hat{u}_k) \leq 0, \quad k \in \{0, \ldots, P\}, v \in \{1, \ldots, n_{fe}\} \tag{2a}$$
$$x^L \leq \hat{x}_{k,v} \leq x^U, \quad k \in \{0, \ldots, P\}, v \in \{1, \ldots, n_{fe}\} \tag{2a}$$
$$u^L \leq \hat{u}_k \leq u^U, \quad k \in \{0, \ldots, M\} \tag{2a}$$
$$\hat{u}_k = \hat{u}_M, \quad k \in \{M+1, \ldots, P\} \tag{2a}$$
$$\mathcal{B}(\hat{x}_{k,v}, \hat{u}_k) = 0, \quad k \in \{0, \ldots, P\}, v \in \{v' : v' \in \{1, \ldots, n_{fe}\}, z_{v'} \in \partial\Omega\} \tag{2a}$$
$$\mathcal{I}(\hat{x}_{k,v}) = 0, \quad k \in \{0\}, v \in \{1, \ldots, n_{fe}\} \tag{2a}$$

where $M \leq P$ is the control horizon, $\eta_{k,v}$ are the coefficients determined by an appropriate quadrature or sampling scheme to approximate the space-time integral [6], and the discretized operator $\hat{\mathcal{F}}_z$ approximates $\mathcal{F}_z$ via finite difference and/or collocation methods to approximate the time and spatial derivatives [6], [7], [8], [9]. In this work, we use a backward finite difference scheme, but other strategies can also be implemented.

This direct transcription approach will typically generate a problem with a large number of variables and nonlinear constraints which in principle can incur an appreciable computational cost. Nevertheless, direct transcription is commonly applied to NMPC problems since efficient solution times are often achieved in practice and path constraints are easily enforced [4], [33], [34], [35], [36]. The literature often refers to direct transcription methods as simultaneous since both $\hat{x}$ and $\hat{u}$ represent decision variables in those problems. Typically, discretized constraints only use a few optimization variables which makes the problem sparse (i.e., the Jacobian and Hessian matrices are mostly comprised of zeros). Most AD routines employed by AMLs exploit this sparsity to efficiently evaluate Jacobian and Hessian matrices [8].

Notably, shooting/sequential methods are a widely studied alternative to solve NMPC problems [15], [28], [29]. In a shooting method, the optimizer optimizes a sequence of control actions, $\{u_k, u_{k+1}, \ldots, u_{k+M-1}\}$, that are given to an embedded integrator that solves the ODEs/PDEs at each iteration [7], [8]. This approach removes the state variables from the optimization problem, reducing its size, but second-order derivatives are rarely computed due to computational cost [8]. We limit the scope of this work to direct transcription since its performance with NMPC problems that embed surrogate models has not been reported in the literature. However, a simple shooting method is included in the benchmark study presented in Section 4 to connect to previous works (e.g., [15]) that used shooting methods.

### 2.3. Replacing PDEs with Surrogate Models

Recent studies have proposed replacing nonlinear ODEs in NMPC problems with NN surrogate models to reduce the computational cost [15], [16], [28], [29]. In [15], Antonelo et al. proposed a framework that uses PINNs as surrogate models for ODEs. In that work, a PINN, based on a fully connected feedforward NN (FNN), is trained to predict the states $\hat{x}$ at any time within the interval $t \in [0, \Delta t]$ given a target time $t$, the current states $\hat{x}_k$, and current control variables $\hat{u}_k$ as inputs. The PINN is used recursively to make predictions over longer time horizons. This methodology was used with NMPC problems solved via shooting methods, resulting in a mild computational cost reduction relative to using the mechanistic ODE model [15], [28], [29]. That approach has not been applied to PDE-constrained NMPC problems, nor problems solved via direct transcription.

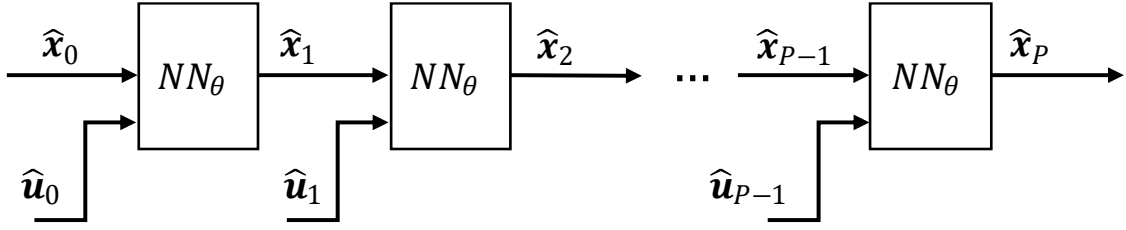

**Figure 1 State propagation over the prediction horizon from the initial state and the control actions at each step in time.**

In our approach, instead of passing $t$ as a continuous input, the physics-informed surrogate model learns a mapping over one time step. Specifically, the NN surrogate maps the current states across all spatial locations $\hat{\boldsymbol{x}}_k \in \mathbb{R}^{n_x \times n_{fe}}$ and current control actions $\hat{\boldsymbol{u}}_k$ to the states at the next sampling step $\hat{\boldsymbol{x}}_{k+1}$. This enables the model to handle varied initial conditions and control inputs without retraining such that it can be propagated over the prediction horizon $P$. Hence, Equation (2a) can be replaced as follows:

$$\hat{\boldsymbol{x}}_{k+1} = NN_\theta(\hat{\boldsymbol{x}}_k, \hat{\boldsymbol{u}}_k), \quad k \in \{0,1,\dots,P-1\}, v \in \{1,\dots,n_{fe}\}. \tag{3}$$

where $NN_\theta$ represents the NN surrogate model with learned parameters $\theta$ that is embedded in the NMPC formulation shown in problem (2). Figure 1 depicts how Equation (3) uses $NN_\theta$ to recursively predict the states over the prediction horizon. Note that a single NN model is considered for the entire time and spatial domain. This work will focus on the use of FNN and CNN architectures for $NN_\theta$ that use physics-informed loss functions as described in Section 3. As mentioned, Eq. (2a) represents general nonlinear path constraints. We directly include the function, $\boldsymbol{g}$, defining these constraints as algebraic constraints in the NMPC problem.

## 3. Surrogate Modelling and Embedding

In this section, we detail the architectures and training procedures for the surrogate $NN_\theta$ models. We also discuss two strategies to embed these models into NMPC problems: ECE and EFE.

### 3.1. PINN Model Structure

Physics-informed NNs based on FNNs (i.e., PINNs) are a common choice because of their simplicity [15], [25], [28], [29], [37]. In the context of PDE surrogates, a key limitation is that FNNs can only accept one-dimensional inputs (i.e., vector inputs). Since $\hat{\boldsymbol{x}}_k$ is a matrix (or a tensor for higher dimensional PDEs), it must be flattened before it is given to the FNN, which removes the spatial relationships inherently in the data structure. The flattened state vector $\bar{\boldsymbol{x}}_k$ is given by:

$$\overline{x}_k = \left[\hat{x}_{1,k,1}, \ldots, \hat{x}_{1,k,n_{fe}}, \hat{x}_{2,k,1}, \ldots, \hat{x}_{2,k,n_{fe}}, \ldots, \hat{x}_{n_x,k,1}, \ldots, \hat{x}_{n_x,k,n_{fe}}\right]^T \tag{4}$$

where $\hat{x}_{c,k,v}$ denotes the $c^{th}$ state variable at time $t_k$ and location $z_v$. This flattened state vector is then concatenated with control variables $\hat{u}_k$ to yield the FNN input vector $r_0$:

$$r_0 = [\overline{x}_k^T \quad \hat{u}_k^T]^T. \tag{5}$$

The output $r_l \in \mathbb{R}^{n_{r,l}}$ ($n_{r,0} = n_x n_{fe} + n_u$) at the $l^{th}$ hidden FNN layer is computed as follows:

$$r_{l+1} = \sigma(E_l r_l + b_l), \quad l \in \{0, \ldots, L-1\} \tag{6}$$

where $E_l \in \mathbb{R}^{n_{r,l+1} \times n_{r,l}}$ denotes the weights matrix for layer $l$, $b_l \in \mathbb{R}^{n_r}$ is the bias vector for layer $l$, $\sigma$ is an activation function that is applied elementwise, and $L$ is the number of hidden layers. For NMPC problems, smooth activation functions (e.g., hyperbolic tangent and sigmoid) are preferred since others functions like the rectified linear unit ($ReLU$) may not be twice differentiable (see Section 3.4). Furthermore, the flattened state variables at the next sampling step $\overline{x}_{k+1}$ are computed as the output of the FNN using $r_L$:

$$\overline{x}_{k+1} = E_L r_L + b_L \tag{7}$$

where $E_L \in \mathbb{R}^{n_x n_{fe} \times n_{r,l}}$ and $b_L \in \mathbb{R}^{n_x n_{fe}}$; $\overline{x}_{k+1}$ can then be reshaped into $\hat{x}_{k+1}$. Figure 2 summarizes the FNN structure.

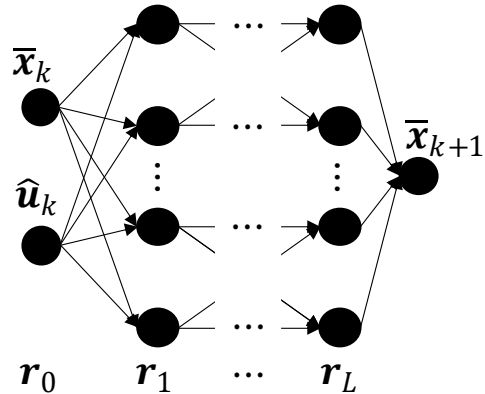

**Figure 2 Depiction of multilayer structure of FNN.**

### 3.2. PICNN Model Structure

Contrary to FNNs, CNNs accept tensor inputs and can take advantage of the spatial relationships in the input data that would otherwise be lost through flattening [38]. This makes CNNs a natural choice for modelling PDEs that are indexed over a spatial domain. CNNs typically accept multiple equally sized tensors as input which are referred to as channels. Thus, for $NN_\theta$, we let each state/control variable at time point $t_k$ correspond to an input channel whose dimension is determined by the size of $\widehat{\Omega}$. Control variables $\widehat{u}_k$ that have no spatial dependency are simply repeated to achieve the correct tensor size. For a 1D PICNN, the collection of input channels $r_0 \in \mathbb{R}^{(n_x+n_u) \times n_{fe}}$ to the first convolutional layer of a PICNN is defined as follows:

$$r_{0,c,v} = \hat{x}_{c,k,v}, \qquad c \in \{1, \dots, n_x\}, v \in \{1, \dots, n_{fe}\} \tag{8}$$

$$r_{0,c+n_x,v} = \hat{u}_{c,k,v}, \qquad c \in \{1, \dots, n_u\}, v \in \{1, \dots, n_{fe}\}. \tag{9}$$

Figure 3 summarizes the multilayer PICNN architecture used in this work. The input tensor $r_0$ is fed through a 1D convolutional (Conv1d) layer with 32 channels that each use a convolutional filter of size 4 and a smooth nonlinear activation function $\sigma$ that is applied elementwise. This operation is repeated over $L$ convolutional layers until $r_L$ is obtained. Note that the hyperparameters and network structure were determined via a trial-and-error until a satisfactory loss function value was obtained. Also, the dimension of each channel is subsequently reduced since no padding is used (OMLT does not currently support padding as discussed in Section 3.4.1). Moreover, pooling layers are excluded for simplicity and because removing them improved the performance of benchmark models used in Section 4.1. Furthermore, the output $r_L$ is flattened, passed through a dense layer with a linear activation function in similar manner to Equation (7), and then reshaped into the final output $\hat{x}_{k+1}$ such that it has the same dimensions as $\hat{x}_k$. A previous study [26] used a similar PICNN structure to predict Darcy flows in a heterogeneous reservoir; control variables were not provided as inputs in that work. A more in-depth discussion of the components used by CNNs is provided in [38].

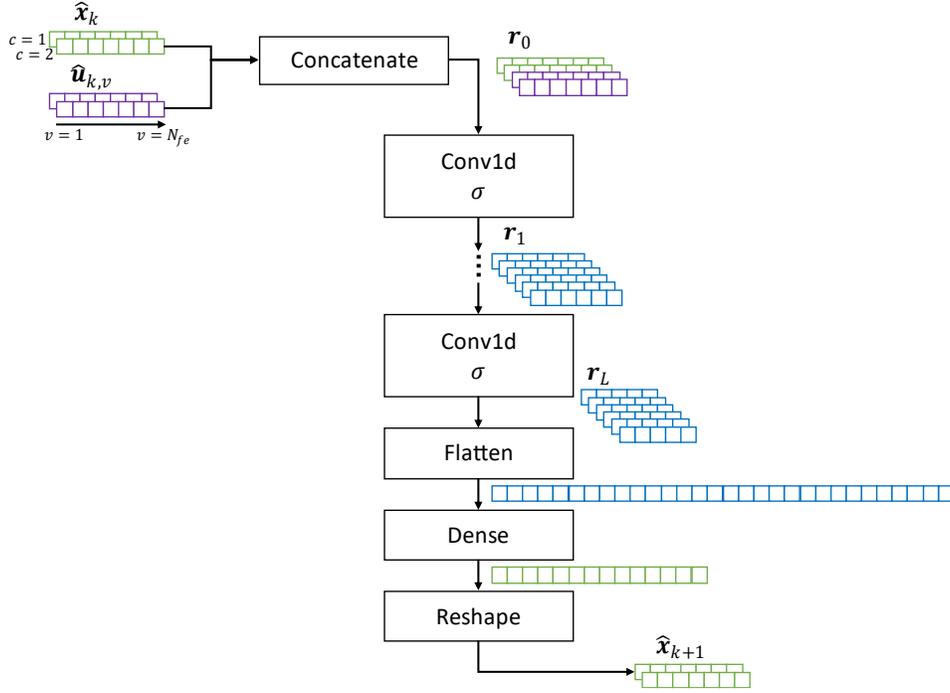

**Figure 3** A schematic of the PICNN architecture. The green blocks represent state variable tensors, the purple blocks represent control variable tensors, and the blue blocks represent the outputs of the internal/hidden layers of the CNN.

### 3.3. Model Training

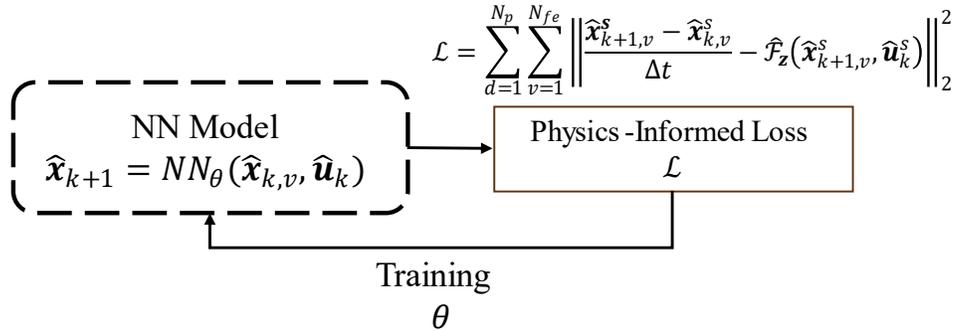

**Figure 4** PINN model training framework.

Both the FNN and CNN surrogate models used in this work are considered physics-informed (i.e., PINNs and PICNNs, respectively) since the original PDEs are incorporated into the loss function used to train each neural network. Specifically, we leverage the transcribed form of the PDE in Equation (2a) to approximate the solution of the PDE where the derivatives are approximated via finite difference. Figure 4 illustrates the training procedure. The input dataset, comprised of the state and control variables at time $t_k$, is generated by collecting uniform random samples of the variables over their feasible sets, $\mathcal{X}$ and $\mathcal{U}$, to obtain $\{(\hat{\boldsymbol{x}}_k^s, \hat{\boldsymbol{u}}_k^s): s \in \{1, \dots, n_s\}\}$ where $n_s$ is the number of samples. The outputs of $NN_\theta$ using this data are treated as predictions of $\hat{\boldsymbol{x}}_{k+1}^s$ which help define the residual $\boldsymbol{R}_v^s \in \mathbb{R}^{n_x}$ relative to the discretized PDE:

$$\boldsymbol{R}_v^s \equiv \frac{\hat{\boldsymbol{x}}_{k+1,v}^s - \hat{\boldsymbol{x}}_{k,v}^s}{\Delta t} - \hat{\mathcal{F}}_z(\hat{\boldsymbol{x}}_{k+1,v}^s, \hat{\boldsymbol{u}}_k^s), \quad v \in \{1, \dots, n_{fe}\}. \tag{10}$$

The sum of squared residuals then determines the loss function $\mathcal{L}$ used to train $NN_\theta$, i.e.,

$$\mathcal{L} = \sum_{s=1}^{n_s} \sum_{v=1}^{n_{fe}} \|\boldsymbol{R}_v^s\|_2^2. \tag{11}$$

A ML environment (e.g., PyTorch) can be used to find optimal model parameters $\theta$ (e.g., the weight matrices and bias vectors) that minimize Equation (11). Gradients for this optimization are provided via backpropagation (a type of AD) which is facilitated by the ML environment [31], [32]. If available, process/historical data can also be added to the loss function, but no such data is used in this work for simplicity.

### 3.4. Embedding Strategies

The trained $NN_\theta$ surrogate model is incorporated into the transcribed PDE-constrained NMPC problem according to Equation (3). Here, the variable bounds in Equations (2a) and (2a) are critical to ensure that the surrogate model not be evaluated outside the domain of its training data. Two strategies are considered here to embed the into the NMPC formulation: (i) directly translating $NN_\theta$ into algebraic constraints (i.e., ECE), and (ii) treating $NN_\theta$ as an external function (i.e., EFE). In addition to these $NN_\theta$ strategies, a third alternative is to use the transcribed PDE constraints (i.e., the mechanistic model) instead of the surrogate model. Figure 5 summarizes these three modelling approaches. The following subsections discuss the details of ECE and EFE.

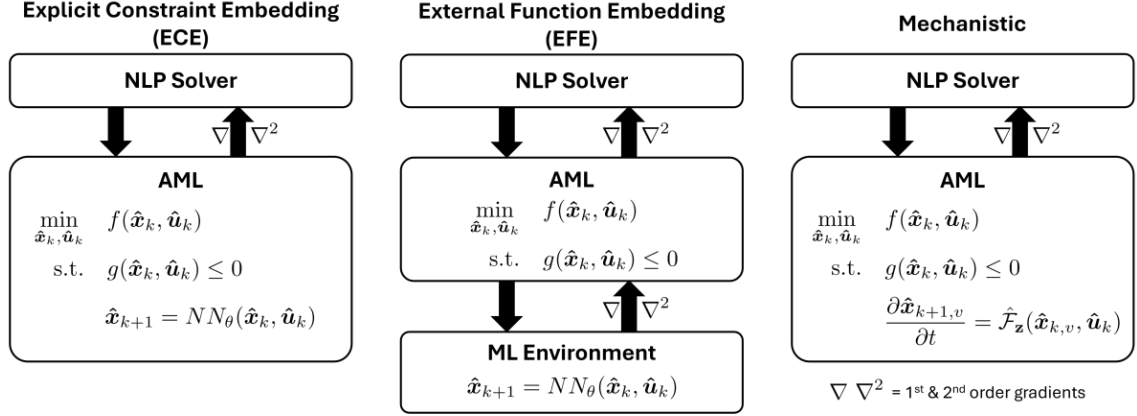

**Figure 5 A summary of the three ways to formulate Problem (2a). A NN surrogate can be embedded via ECE or EFE, or the transcribed PDE can be used directly instead.**

*3.4.1. Explicit Constraint Embedding (ECE)*

The literature presents several algebraic representations of FNNs that can be cast as constraints in AMLs as part of a larger optimization model. The two main classes of representations are full-space (FS) and reduced-space (RS) representations. For the $l^{\text{th}}$ FNN layer with a smooth nonlinear activation function that follows Equation (6), the FS representation is as follows:

$$r'_{l+1,p} = \sum_{p'=1}^{n_r} E_{l,p,p'} r_{l,p'} + b_{l,p}, \quad l \in \{0, \ldots, L-1\}, p \in \{1, \ldots, n_r\} \tag{12}$$

$$r_{l+1,p} = \sigma(r'_{l+1,p}), \quad l \in \{0, \ldots, L-1\}, p \in \{1, \ldots, n_r\} \tag{13}$$

where $r_{l,p}$ is the $p^{\text{th}}$ element of $\boldsymbol{r}_l$, $r'_{l,p}$ is $r_{l,p}$ before activation, $E_{l,p,p'}$ is the $(p, p')$ entry of $\boldsymbol{E}_l$, $b_{l,p}$ is the $p^{\text{th}}$ entry of $\boldsymbol{b}_l$ [30], [39]. Thus, each FNN layer generates $2n_r$ additional constraints and variables to the optimization problem. RS representations avoid adding additional constraints and variables for intermediate NN layers by recursively substituting the outputs of one layer into the inputs of the next layer; however, this reduction in the number of constraints/variables produces constraints with much larger expressions that may or may not make the optimization formulation more effective. Both methods were tested in this work, we refer to FS and RS embedding methods as ECE-FS and ECE-RS, respectively.

Convolutional layers in CNNs use specialized convolutional filters to extract features from tensor data. The FS representation of a 1D convolutional layer with a smooth activation function is:

$$r'_{l+1,c,v} = \sum_{c'=1}^{n_c} \sum_{v'=1}^{n_\kappa} \kappa_{l,c,c',v'} r_{l,c',v+v'-1} + b_{l,c} \tag{14}$$

$$r_{l+1,c,v} = \sigma(r'_{l+1,c,v}) \tag{15}$$

for $l \in \{0, \ldots, L-1\}, c \in \{1, \ldots, n_c\}, v \in \{1, \ldots, n_{v,l+1}\}$, where $n_c$ is the number of channels at each layer, $n_\kappa$ is the size of the convolutional filter, and $n_{v,l+1} = n_{v,l} - n_\kappa + 1$ is tensor length at layer $l+1$ (no padding is used with a stride of one) [38]. The convolutional filter (i.e., kernel) for the input channel $c'$ to output channel $c$ in layer $l$ is represented by $\kappa_{l,c,c'} \in \mathbb{R}^{n_\kappa}$. Hence, each layer adds an additional $2n_c \times n_{v,l+1}$ constraints and variables to the optimization problem.

It the context of NMPC, it is important to use NN surrogate models with smooth activation functions that are twice differentiable to produce formulations that are well-suited for NLP solvers such as Ipopt. Non-smooth activation functions, such as ReLU, can be reformulated using big-M constraints [40] or partition-based formulations [41] that introduce integer variables thus converting the NLP into a mixed-integer NLP, which would likely be computationally expensive for online NMPC calculations. NNs with ReLU activation functions can also be reformulated via complementary constraints to avoid introducing integer variables; however, NLP solvers often struggle with complementarity constraints due to degeneracy [42], [43].

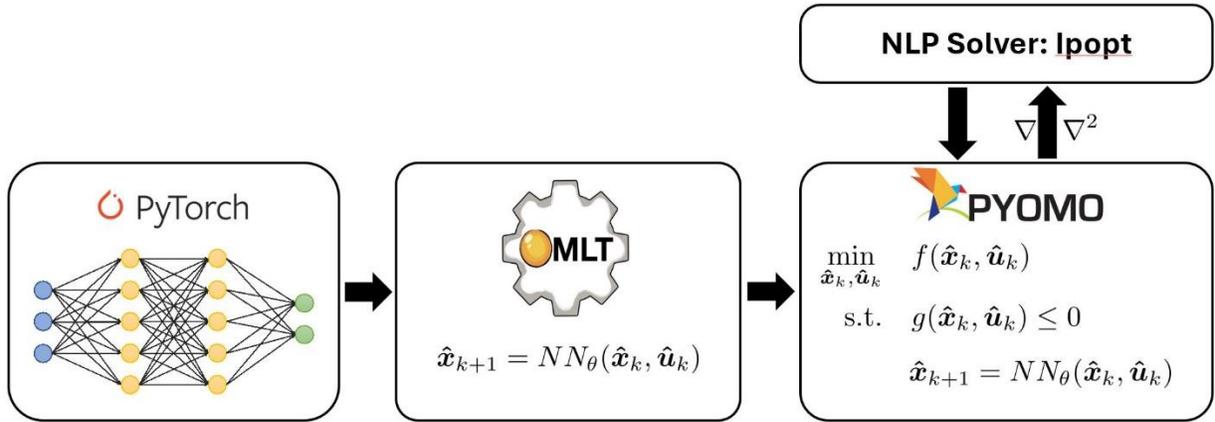

**Figure 6 Illustration of using OMLT to embed $NN_\theta$ in a Pyomo model via ECE.**

Several software packages have been developed to embed trained NNs into optimization problems via ECE methods [44] such as MeLOn [45], JANOS [46], OptiCL [47], OMLT [30], and PySCIPOpt-ML [48]. Of these, we select OMLT since it is the only software tool to support FNNs and CNNs with smooth nonlinear activation functions. Also, OMLT is a Python package that automates the application of ECE for Pyomo models via an extensive library of ECE-FS and ECE-RS methods for a wide variety of ML models [30]. For trained FNNs and CNNs from PyTorch [31] or Tensorflow [32], OMLT creates the appropriate Pyomo constraints and variables that are added to the overarching optimization model object as shown in Figure 6. While OMLT supports FS and RS representations for FNNs, it currently only supports FS representation for CNNs that do not use padding. These limitations played a part in constraining the CNN architecture presented in Section 3.2.

*3.4.2. External Function Embedding (EFE)*

As an alternative to ECE, EFE defines the surrogate model as an external function in the AML such that it is evaluated and differentiated externally in the ML environment. Specifically, we seek an external function $\boldsymbol{w}: \mathbb{R}^{n_y} \to \mathbb{R}^{n_x*P}$ that enforces Equation (3) as follows:

$$\boldsymbol{w}(\boldsymbol{y}) = \begin{bmatrix} \hat{\boldsymbol{x}}_1 - NN_\theta(\hat{\boldsymbol{x}}_0, \hat{\boldsymbol{u}}_0) \\ \vdots \\ \hat{\boldsymbol{x}}_P - NN_\theta(\hat{\boldsymbol{x}}_{P-1}, \hat{\boldsymbol{u}}_{P-1}) \end{bmatrix} = 0 \qquad (16)$$

where $\boldsymbol{y} \in \mathbb{R}^{n_y}$ includes all the transcribed state and control variables:

$$\boldsymbol{y} = \left[\hat{\boldsymbol{x}}_0^T, \dots, \hat{\boldsymbol{x}}_P^T, \hat{\boldsymbol{u}}_0^T, \dots, \hat{\boldsymbol{u}}_{P-1}^T\right]^T. \qquad (17)$$

The AML then requires routines to evaluate $\boldsymbol{w}(\boldsymbol{y})$, i.e., the Jacobian $\frac{\partial}{\partial \boldsymbol{y}}\boldsymbol{w}(\boldsymbol{y})$, and the Hessian $\frac{\partial^2}{\partial \boldsymbol{y}^2}\boldsymbol{\lambda}^T \boldsymbol{w}(\boldsymbol{y})$. $\boldsymbol{\lambda} \in \mathbb{R}^{n_x*P}$ are Lagrange multipliers that correspond to each element of $\boldsymbol{w}(\boldsymbol{y})$ being enforced as an equality constraint. The Lagrange multipliers assign the contribution of each constraint to the gradient of the optimization problem at optimality conditions. The Lagrange multipliers are computed by the optimizer and provided to the ML environment to compute the Hessian. The ML environment used to train $NN_\theta$ can readily use its AD capabilities to provide the function, Jacobian, and Hessian evaluations at the current value of $\boldsymbol{y}$ provided by the optimization solver through the AML's external function interface. Therefore, direct transcription is implemented since the solver has complete control over the optimization variables $\boldsymbol{y}$. One potential drawback to the EFE approach is that the communication between the AML and the ML environment along with any needed data type conversions (e.g., converting between 32-bit and 64-bit floating point numbers) may become a performance bottleneck.

In this work, we select PyTorch as the ML environment since it is one of most popular. To efficiently evaluate each constraint in Equation (16), all the corresponding inputs can be fed as a batch to the NN. PyTorch provides a vectorized approach via `torch.func`. It provides the functions `jacrev` and `jacfwd` that evaluate the Jacobians and the function `hessian` to evaluate the Hessians. Here, `jacrev` evaluates the Jacobian row by row and `jacfwd` column by column, whereas `hessian` applies `jacrev` followed by `jacfwd`. Since $n_y > P$, the Jacobian $\frac{\partial}{\partial y}w(y)$ has fewer rows than columns such that `jacrev` should be more efficient than `jacfwd`.

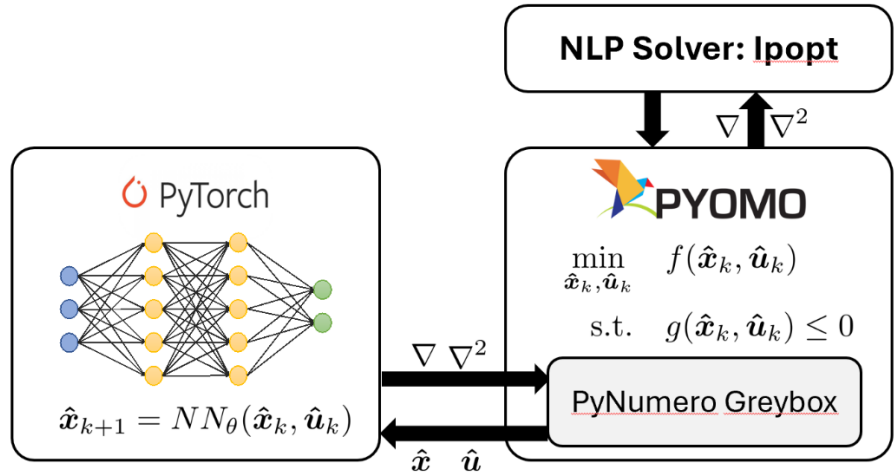

**Figure 7 Schematic of the EFE approach using PyNumero and PyTorch.**

With the evaluation routines for Equation (16) set up in PyTorch, we leverage PyNumero's `ExternalGreyBoxModel` interface to embed these as constraints in Pyomo, which are evaluated externally in PyTorch [49]. Figure 7 depicts the interaction between the PyNumero interface and the external PyTorch model. As shown in this Figure, PyNumero passes the current variable values from the NLP solver to the external model which in turn evaluates $w(y)$, $\frac{\partial}{\partial y}w(y)$, and $\frac{\partial^2}{\partial y^2}\lambda^T w$ (signified by $\nabla$ and $\nabla^2$ in the figure). PyNumero then passes these evaluated vectors/matrices to the NLP solver which then updates the variable values. This procedure is repeated each time the solver requires a function/gradient evaluation.

*3.4.3. Automatic Differentiation*

The discussed embedding strategies fundamentally differ in the ways they perform AD. In general, AD refers to the evaluation of exact derivatives via symbolic derivative rules enforced on function composition of elementary operators such as product, summation, exp, and sin [50], [51]. Note that AD is not symbolic differentiation, which applies the rules of differentiation to generate a symbolic derivative expression based on an input symbolic expression. Instead, AD tracks intermediate variables and their derivatives in a sequence of operations to return an evaluator of exact derivatives. AD is available in forward and reverse modes. Forward mode AD accumulates the gradients from the input variables to the outputs by applying the chain rule. Reverse mode AD (a common choice in AMLs) propagates the derivatives backwards from a given output to the inputs. Backpropagation is a special case of reverse mode AD applied to neural networks, though it is also commonly referred to as AD in the ML community [51].

Both AMLs and ML environments often integrate specialized AD frameworks for evaluating gradients. Here, AMLs typically use a reverse mode AD that is tailored to exploit the sparsity frequently present in NLPs. While ML environments use AD routines that have been developed to efficiently train deep learning ML models (e.g., backpropagation). For embedding NN surrogates into NMPC problems, it is not obvious which approach will more efficiently provide gradients to an NLP solver. The ECE method leverages the AD provided by the AML, while EFE makes use of the AD from the ML environment. The next section benchmarks these different surrogate model embedding approaches in the context of PDE-constrained NMPC.

## 4. Closed-Loop NMPC Benchmarks

This section details benchmark studies to compare the performance of ECE-FS, ECE-RS, and EFE for embedding PINN and PICNN PDE surrogates in closed-loop NMPC. All the computational results are collected on a Windows 10 machine equipped with 32 GB RAM and an Intel® Core™ i9-10980HK CPU @ 2.40 GHz. PyTorch 2.2.1, OMLT 1.1, Pyomo 6.5.0, PyNumero 1.3, and IPOPT 3.14.9 are used on Python 3.8. All the source code is available at https://git.uwaterloo.ca/ricardez_group/benchmarking-nn-surrogate-embedding-methods-in-nmpc.git.

### 4.1. Models and Surrogate Training

The benchmarks are based on three PFR models on increasing complexity: a 1D isothermal PFR, a non-isothermal 1D PFR, and a highly nonlinear methane steam reforming PFR. Note that all the NN surrogates are trained using the ADAM optimizer with an initial learning rate of 0.01 and an exponential decay factor of 0.7 every 100 epochs on PyTorch 2.0.1, with 100,000 samples. Each of the PFR case studies is described next.

*4.1.1. Benchmark 1: Isothermal Plug-Flow Reactor*

This model captures the concentration $C \in [0.1, 1]$ of a reactant through a PFR over time $t \in \mathcal{T}$ and reactor volume $\mathbf{z} \in \Omega = [0, 1]$ via the molar balance PDE:

$$\frac{\partial C}{\partial t} = -F\frac{\partial C}{\partial \mathbf{z}} - k_{rxn}C^2, \quad t \in \mathcal{T}, \mathbf{z} \in \Omega \tag{18}$$

where $k_{rxn} = 1$ is the reaction rate constant, and $F \in [0, 1]$ is the volumetric flow rate. The boundary and initial conditions are as follows:

$$C(t, 0) = C_{in}(t), \quad t \in \mathcal{T} \tag{19}$$
$$C(t_0, \mathbf{z}) = C_0(\mathbf{z}), \quad \mathbf{z} \in \Omega \tag{20}$$

where $C_{in} \in [0.1, 1]$ is the inlet concentration and $C_0$ is the initial concentration profile over the reactor volume. The NMPC controller seeks to have $C_{out}(t) = C(t, 1)$ track a setpoint by manipulating $C_{in}$ and $F$. Hence, $\mathbf{x} = C$ and $\mathbf{u} = [C_{in}\ F]^T$. For direct transcription, we use a sampling time $\Delta t = 0.1\ s$ and discretize $\mathbf{z}$ with a uniform grid of $n_{fe} = 10$ finite elements, which were chosen from preliminary simulation tests.

A PINN surrogate model with six hidden layers that each use 24 neurons is trained using the structure and procedure described in Section 3. This model has 12 inputs (10 transcribed states and 2 control variables) and 10 outputs corresponding to the transcribed states. A PICNN with 2 convolutional layers is also trained following the structure and procedure described in Section 3. This PICNN uses 3 input channels (1 state and 2 control variables) and 1 output channel (1 state variable), where each channel has a size of 10 since $n_{fe} = 10$. Note that the focus of this work is not the accuracy of the ML models but rather on the embedding methods. Hence, we applied a trial-and-error approach to determine the network's hyperparameters. Figure B.1 in the supplementary material shows the loss function values evaluated for each epoch during the training of the PINN and the PICNN; both converge to reasonably low loss values. While it is difficult to verify that the models accurately capture all possible state and control variable combinations, we observe that increasing the size of the input dataset does not further improve accuracy. Additionally, the training data and test data achieve similar losses in Figure B.1 which suggests that the surrogate models are generalizing to data points not observed during training and that they do not overfit. Once trained, the PINN and PICNN models require 0.000421s and 0.000436s on average for evaluation, respectively, which provides a modest speed-up over direct simulation of the PDE model that requires 0.000504s on average.

*4.1.2. Benchmark 2: Non-isothermal Plug Flow Reactor with Heat Exchange*

This model is adapted from [52]. As in Benchmark 1, this PFR model uses Equations (18)-(20) to model the concentration where $C \in [0.2, 1.9]$, $\Omega = [0, 5000]$, and $F \in [1.062, 4.249]$. The PFR model also incorporates an energy balance PDE:

$$\rho C_p \frac{\partial T}{\partial t} = -\rho C_p F \frac{\partial T}{\partial z} + Ua(T_a - T) - \Delta H k_{rxn} C^2, \quad t \in \mathcal{T}, z \in \Omega \tag{21}$$

where $T: (\mathcal{T} \times \Omega) \mapsto [300, 320]$ is the reactor temperature, $\rho C_p = 295.7$ is the fluid heat capacity, $Ua = 1.389$ is the convective heat transfer coefficient, $\Delta H = -34.5$ is the heat of reaction, and $T_a \in [300, 320]$ is the exchanger heating temperature. The additional initial/boundary conditions are:

$$T(t, 0) = T_{in}(t), \quad t \in \mathcal{T} \tag{22}$$

$$T(t_0, z) = T_0(z), \quad z \in \Omega \tag{23}$$

where $T_{in}(t) \in [300, 320]$ is the inlet temperature and $T_0$ is the initial temperature profile in the reactor. Moreover, the Arrhenius equation models the temperature dependence of the reaction rate constant $k_{rxn}$:

$$k_{rxn} = k_0 \exp\left(-\frac{E_A}{RT}\right), \quad t \in \mathcal{T}, z \in \Omega \tag{24}$$

where $k_0 = 4.79 \cdot 10^7$ is the preexponential factor, $E_A = 65730$ is the activation energy, and $R = 8.314$ is the ideal gas constant. Here, state variables are defined $x = [C, T]^T$ and the control variables are $\mathbf{u} = [C_{in}, F, T_a, T_{in}]^T$. The model is transcribed with a sampling time of $\Delta t = 0.1\ s$ and again $n_{fe} = 10$ uniform finite elements are used.

We were unable to find a PINN surrogate that was able learn the PDEs with sufficient accuracy such that the NMPC formulation could find a feasible solution. This can likely be attributed to the loss of information that occurs with the flattening step discussed in Section 3.1. We tested different PINN architectures, but the problem persisted. There may exist a PINN architecture that overcomes this challenge; finding such a PINN is beyond the scope of this work. A PICNN, which does not require input flattening, is successfully able to learn the PDEs using the same structure of the Benchmark 1 PICNN, adjusting it to appropriately use 6 input and 2 output channels. Figure B.2 in the supplementary material shows the PICNN training loss curves which show a similar level of accuracy as the PICNN trained for Benchmark 1. On average, the PICNN requires 0.000253s to evaluate, making it an order-of-magnitude faster than direct simulation of the PDE model which requires 0.00316s.

*4.1.3. Benchmark 3: Steam Reformer PFR*

This model considers an isobaric 1D PFR that undergoes a multi-component gas-phase reaction that is assumed to behave as an ideal gas. The governing chemical reactions are:

$$CH_4 + H_2O \rightleftharpoons CO + 3H_2 \tag{25}$$
$$H_2O + CO \rightleftharpoons H_2 + CO_2 \tag{26}$$
$$CH_4 + 2H_2O \rightleftharpoons CO_2 + 4H_2 \tag{27}$$

The concentrations $C_s$ of each component $s \in \{CH_4, H_2O, H_2, CO_2, CO\}$ are determined by a system of molar balance PDEs:

$$\frac{\partial C_s}{\partial t} = -\frac{1}{A}\frac{\partial F_s}{\partial \mathbf{z}} + \rho_c \sum_{j=1}^{3} \nu_{js} RR'_j, \quad s \in \{CH_4, H_2O, H_2, CO_2, CO\}, t \in \mathcal{T}, \mathbf{z} \in \Omega \tag{28}$$

where $F_s \in \mathbb{R}_+$ is the molar flow rate, $A \in \mathbb{R}_+$ is the reactor cross-sectional area, $\mathbf{z} \in \Omega$ is the position along the length of the reactor, $\rho_c \in \mathbb{R}_+$ is the catalyst packed density, $\nu_{js} \in \mathbb{R}$ is the stochiometric coefficient of the component $s$ in the $j^{\text{th}}$ reaction, and $RR'_j \in \mathbb{R}$ is the rate of the $j^{\text{th}}$ reaction. The component concentrations and flowrates are described as follows:

$$y_s = \frac{F_s}{\sum_s F_s}, \quad s \in \{CH_4, H_2O, H_2, CO_2, CO\}, t \in \mathcal{T}, \mathbf{z} \in \Omega \tag{29}$$
$$P_s = y_s P_{tot}, \quad s \in \{CH_4, H_2O, H_2, CO_2, CO\}, t \in \mathcal{T}, \mathbf{z} \in \Omega \tag{30}$$
$$C_s = \frac{P_s}{RT}, \quad s \in \{CH_4, H_2O, H_2, CO_2, CO\}, t \in \mathcal{T}, \mathbf{z} \in \Omega \tag{31}$$

where $y_s \in [0, 1]$ is the component molar fraction, $P_s \in \mathbb{R}_+$ is the component partial pressure, and $P_{tot} \in \mathbb{R}_+$ is the total reactor pressure. The reaction rates $RR'_j$ are highly nonlinear functions of the reactant partial pressures and the reactor temperature. The auxiliary equations for these along with all the model parameters are provided in Section A of the supplementary materials. These values are derived using the kinetic parameters from [53] and the thermodynamic parameters from [54]. The initial/boundary conditions are:

$$F_s(t, 0) = F_{s,in}(t), \quad s \in \{CH_4, H_2O, H_2, CO_2, CO\}, t \in \mathcal{T} \tag{32}$$
$$F_s(t_0, \mathbf{z}) = F_{s,0}(\mathbf{z}), \quad s \in \{CH_4, H_2O, H_2, CO_2, CO\}, \mathbf{z} \in \Omega \tag{33}$$

where $F_{s,in} \in \mathbb{R}_+$ is the inlet flowrate of each component and $F_{s,0}$ is the initial component flow profile. The state variables are $\mathbf{x} = [F_{CH_4} \ F_{H_2O} \ F_{H_2} \ F_{CO_2} \ F_{CO}]^T$ and the control variables are $\mathbf{u} = [F_{H_2,in} \ F_{CH_4,in}]^T$. The setpoints are set with respect to the outlet flowrates $F_{H_2,out}$, $F_{CH_4,out}$ and $F_{CO,out}$. The model is transcribed using $\Delta t = 0.004\,s$ and $n_{fe} = 50$ uniformly spaced finite elements.

As in the previous case study, we are unable to find a PINN that is sufficiently accurate to allow the NMPC problem to find a feasible solution. However, a PICNN with 3 hidden convolutional layers and the appropriate number of input/output channels can be trained to a sufficient accuracy for the purposes of this benchmarking study. Figure B.3 in the supplementary material shows the training loss curve for this PICNN. The final evaluated loss is relatively large, leading to some model mismatch relative to directly simulating the PDEs. While such a PICNN would not be a viable surrogate model in practice, it is sufficient for the purposes of benchmarking candidate surrogate model embedding strategies which is the focus of this work. This PICNN requires 0.00162s on average to evaluate, making it two orders-of-magnitude faster than the direct simulation of the mechanistic process model, which requires 0.132s on average.

### 4.2. Numerical Results

This section describes the methodology used to evaluate all the NMPC benchmarks, presents the numerical results, and discusses the key findings.

#### 4.2.1. NMPC Benchmarking Methodology

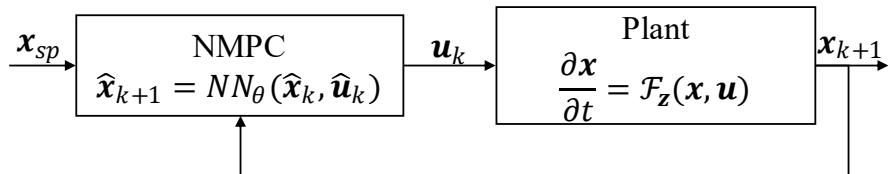

**Figure 8 Closed-loop NMPC framework where the NMPC controller embeds a surrogate model.**

Closed-loop NMPC is simulated on each benchmark model using the candidate surrogate embedding approaches. Figure 8 outlines the simulation framework. An NMPC controller, implemented in Pyomo that uses the same transcription points described in Section 4.1, tracks the step change in $x_{sp}$ as defined in Table 1 and gives the control action $\hat{u}_k$ at each sampling step to the plant model. The plant model (comprised of the PDEs described in Section 4.1) for benchmarks 1 and 2 is simulated via the method of lines, where each PDE is discretized over the spatial domain via finite differences to generate a set of dynamic ODEs which are integrated via Scipy's `solve_ivp` using the Runge-Kutta method of order 5(4) [55], [56]. The spatial discretization was a backward finite difference with the same step sizes used by the PINN/PICNN models (see Section 4.1). For Benchmark 3, the plant was the mechanistic Pyomo model discretized in both time and spatial domains using the same step size of the PINN/PICNN models. The objective function that we employed for all benchmark problems can generally be described by equation (34), where $y_{m,k}$ is the $m$th controlled variable and $u_{l,k}$ the $l$th manipulated variable at time $k$. $L_m$ and $W_l$ are the corresponding weights. Parameters for the NMPC controller and the plant simulator are provided in Table 2, including the weights for Objective (34). The simulation horizons are 100, 150 and 50 for Benchmarks 1, 2 and 3, respectively. For each simulation, the NMPC controller is first initialized with a simulated feasible path for given initial state constant control action over the prediction horizon. It is warm-started on subsequent solutions using the previous solution.

$$\min \sum_{k=0}^{P} \left( \sum_{m=1}^{n_y} L_m (y_{m,sp} - y_{m,k})^2 \right) + \sum_{k=0}^{P-1} \left( \sum_{l=1}^{n_u} W_l \Delta u_{l,k}^2 \right) \tag{34}$$

**Table 1 Setpoint changes used in closed-loop simulations.**

| Benchmark | Controlled-Variable ($y_{m,k}$) | Setpoint Change Time Step $k$ | Initial Setpoint ($y_{m,sp}$) | Final Setpoint ($y_{m,sp}$) |
|---|---|---|---|---|
| 1 | $C_{out}$ | 50 | 0.4 | 0.3 |
| 2 | $C_{out}$ | 100 | 570 | 760 |
| 2 | $T_{out}$ | 50 | 315 | 312 |
| 3 | $F_{H_2,out}$ | 15 | 5.69 | 6.69 |

**Table 2 Summary of objective function weights, control and prediction horizons by benchmark.**

| Benchmark | Controlled Variables ($y_{m,k}$) | | Manipulated Variables ($u_{l,k}$) | |
|---|---|---|---|---|
| | Variable | Weight ($L_m$) | Variable | Weight ($W_l$) |
| 1 ($M = 10$, $P = 40$) | $C_{out}$ | 1 | $F$ | 1 |
| | -- | -- | $C_{in}$ | 1 |
| 2 ($M = 10$, $P = 40$) | $C_{out}$ | 2.77E-07 | $F$ | 5.19E+04 |
| | $T_{out}$ | 2.50E-03 | $C_{in}$ | 2.77E-07 |
| | -- | -- | $T_a$ | 2.50E-03 |
| | -- | -- | $T_{in}$ | 2.50E-03 |
| 3 ($M = 10$, $P = 30$) | $F_{CH_4,out}$ | 3.79E-01 | $F_{CH_4,in}$ | 3.21E-01 |
| | $F_{H_2,out}$ | 1.76E-01 | $F_{H_2O,in}$ | 1.07E-01 |
| | $F_{CO,out}$ | 1.17E+01 | $T_{in}$ | 1.18E-03 |

As a baseline, the closed-loop simulation is carried out on each benchmark model using the mechanistic PDE model according to Problem (2a) via Pyomo.dae. Then the ECE-FS, ECE-RS, and EFE embedding strategies are tested as applicable for each of the surrogate models described in Section 4.1. Note that the EFE methods evaluate the surrogate models on CPU instead of GPU since data transfer between GPU and CPU memory is prohibitively slow. While the focus of this study is on NMPC problems solved via direction transcription, for comparison purposes we also implemented a simple shooting method on Benchmark 1 similar to the approach used in [15]. This method uses the SciPy `fmin_slsqp` optimizer that implements a sequential least-squares quadratic programming approach [56]. The mechanistic models are integrated in the same way the plant model is integrated whereas the surrogate models are evaluated directly with PyTorch. More advanced shooting approaches (e.g., multiple-shooting [9]) might perform better, but the implemented approach is used for illustrative purposes and because it is comparable to the approach used in [15].

*4.2.2. Wall-Clock Timing Results*

Table 3 summarizes the wall-clock times incurred for each NMPC problem solution at each step of the closed-loop simulations and the number of variables in the NMPC problem. Even though the surrogate models evaluate faster on average, the mechanistic NMPC problem is significantly more performant on all benchmarks, highlighting that replacing a nonlinear PDE model with a NN surrogate is not always advantageous in the context of NMPC. For ECE, this can be explained by the fact that the surrogate models are larger (in terms of variables) than the mechanistic model. This also explains why the PINN outperforms the PICNN in Benchmark 1 where the ECE-FS approach uses 55.4% less variables with the PINN. In the case of EFE, the decrease in performance is likely explained by the AML providing more efficient AD on the mechanistic model than PyTorch can provide for the surrogate.

**Table 3 Summary of wall-clock times and number of variables reported for the solution each NMPC problem solved in each of the candidate modelling approaches.**

| Benchmark | Solution Approach | Model | Ave. Wall-Clock Time (s) | 1$^{st}$ Step Wall-Clock Time (s) | # of Variables |
|---|---|---|---|---|---|
| 1 | ECE-FS | PINN | 1.180 | 12.299 | 15120 |
| | ECE-RS | PINN | 1.332 | 2.216 | 2320 |
| | EFE | PINN | 0.402 | 1.890 | 430 |
| | Shooting | PINN | 4.09 | 13.704 | 20 |
| | ECE-FS | PICNN | 5.382 | 48.361 | 33920 |
| | EFE | PICNN | 1.427 | 4.377 | 430 |
| | Shooting | PICNN | 5.040 | 13.260 | 20 |
| | Shooting | Mechanistic | 1.284 | 9.342 | 20 |
| | Direct Transcription | Mechanistic | 0.120 | 0.273 | 1461 |
| 2 | ECE-FS | PICNN | 20.047 | 425.341 | 39680 |
| | EFE | PICNN | 9.586 | 35.356 | 860 |
| | Direct Transcription | Mechanistic | 0.467 | 2.107 | 2922 |
| 3 | ECE-FS | PICNN | 1995.83 | 25805.03 | 327120 |
| | EFE | PICNN | 2149.671 | 1761.363 | 7780 |
| | Direct Transcription | Mechanistic | 16.364 | 73.147 | 75527 |

When a surrogate model is used, the EFE approach almost always exhibits superior performance relative to the ECE methods, which can likely be attributed in part to the EFE approach not introducing any auxiliary variables/constraints that increase the NMPC problem size. For instance, in Benchmark 3, the EFE formulations use 7,780 variables instead of 327,120 for ECE-FS, as shown in Table 3. The effect of the auxiliary variables introduced by the ECE methods is particularly pronounced on the initial NMPC solution of each simulation, where significantly larger wall-clock times are reported due to the difficulty in finding good initial guesses for the auxiliary variables. Notably, ECE-FS outperforms the EFE approach in Benchmark 3 by 7.2% on average, but its clock-time on the initial solution is 1,365% slower. The ECE-RS approach (which is limited to the PINN) does significantly improve the initial solution time relative to ECE-FS (making it 4.6 times smaller); a similar performance boost would be likely observed with ECE-RS on the PICNN models, but this is not currently supported by OMLT.

Figure 9 plots the wall-clock times achieved with ECE-FS and EFE for embedding the PICNN of Benchmark 1 against wall-clock times achieved with the transcribed mechanistic PDE model. This further illustrates how the extent to which the initial solution time exceeds subsequent solution times is significantly larger for ECE-FS than the other approaches. It also shows that, on each step of the closed-loop simulation, EFE outperforms ECE-FS; and both are outperformed by the mechanistic model. Notably, the wall-clock time increases by a similar amount in all cases at the setpoint change as expected. We observe similar patterns in the other benchmark models, with the exception that the warm-started ECE-FS wall-clock times are slightly better those of EFE in Benchmark 3.

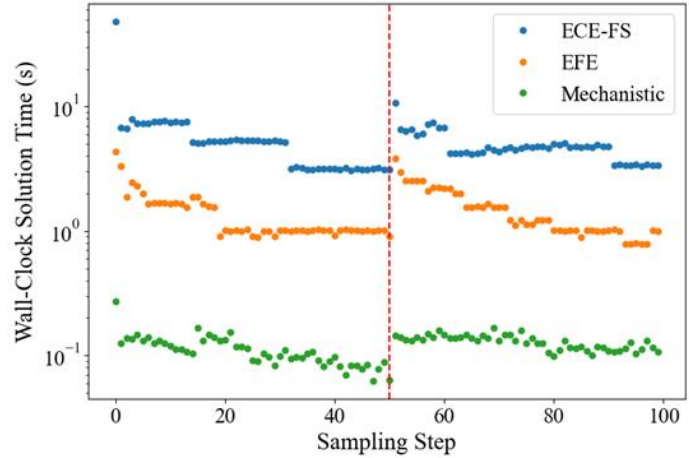

**Figure 9 NMPC wall-clock times for Benchmark 1 at each sampling step for ECE-FS and EFE using a PICNN surrogate in comparison to the transcribed mechanistic baseline. The red dashed line denotes the time of the setpoint change.**

The shooting methods similarly showed that the surrogate models are not able to outperform the mechanistic PDE models considered in this work. This stands in contrast to the results presented by [15] which showed that a PINN surrogate reduced the solution time by 2.6%-23.3% on the reported ODE case studies. However, the PDE models in this work required the use of larger NN surrogate models to capture the high-dimensional behaviour of the PDEs. Even though shooting drastically reduced the number of decision variables, all the direct transcription approaches achieved significantly better wall-clock times than their shooting counterparts, i.e., the mechanistic and PINN results are an order-of-magnitude faster. One exception is that shooting slightly outperforms ECE-FS for the PICNN but is still 2.5 times slower than the EFE approach. We note that more sophisticated shooting methods have been developed [9], [57], [58], [59], [60] which might improve performance, but their implementation is beyond the scope of this work. In this case, shooting is likely less performant because the PDEs are simulated at each solver iteration and the `fmin_slsqp` solver estimates the Jacobian via numerical methods.

*4.2.3. Solution Consistency Verification*

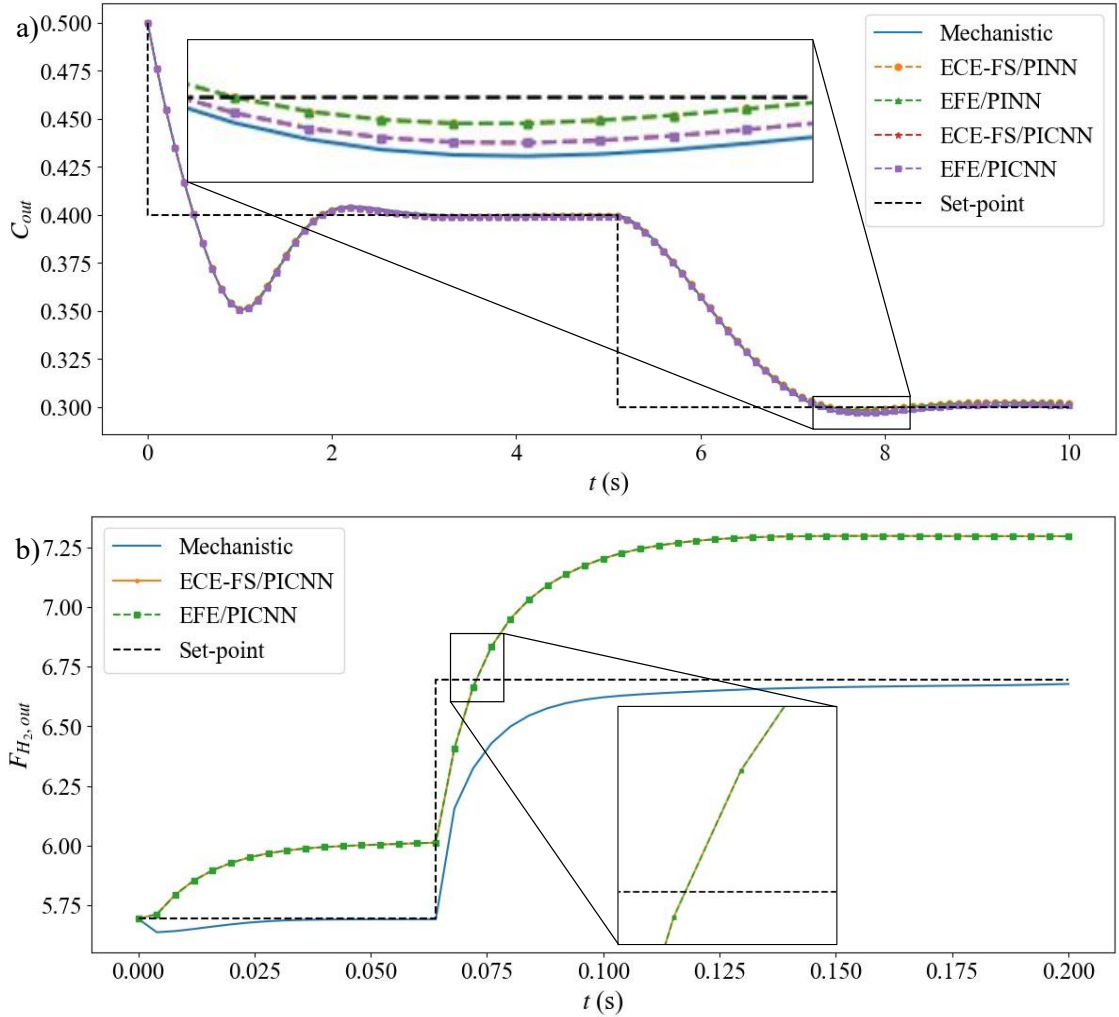

**Figure 10 NMPC setpoint tracking performance for Benchmarks 1 and 3, a) and b), respectively, for selected solution approaches. Benchmark 1 shows the outlet concentration from the PFR. Benchmark 3 shows the outlet flow rate of hydrogen from the steam reformer.**

Looking further into the simulation results, we examine the closed-loop trajectories of the state variables for consistency. Figure 10 shows the closed-loop setpoint tracking results for Benchmarks 1 and 3 collected using ECE-FS, EFE, and the transcribed mechanistic PDE model. ECE-RS is omitted since, it achieves the same trajectories as its ECE-FS counterpart. These trajectories verify that the trajectories are indistinguishable from one another when the same model is used regardless of the embedding strategy. As expected, the trajectories generated with the PINN and PICNN trajectories are nearly identical to those of the mechanistic model for Benchmarks 1 and 2 (omitted for conciseness in presentation) since these surrogates achieved a low residual during training. The surrogate model is not able to match the mechanistic results in Benchmark 3 since it was not sufficiently accurate as discussed in Section 4.1.3. However, the embedded methods still return the same trajectories such that they can be fairly compared.

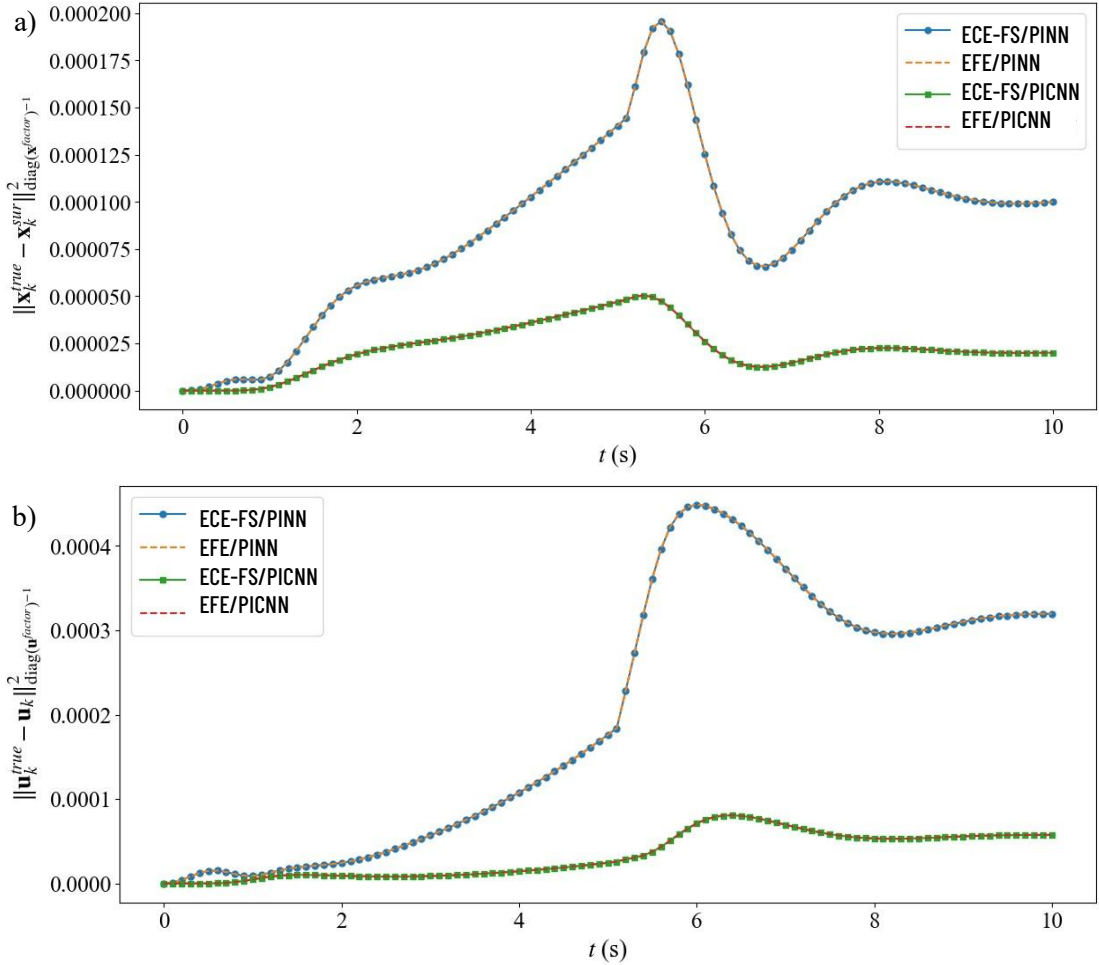

**Figure 11 State trajectory error, a), and control trajectory error, b), plotted against time for Benchmark 1.**

To further confirm consistency in the state and control trajectories obtained at each NMPC solution for each model, we used a distance metric normalized by the mechanistic solution, i.e.,

$$\|\widehat{\boldsymbol{x}}^{true} - \widehat{\boldsymbol{x}}^{sur}\|^2_{\text{diag}(\widehat{\boldsymbol{x}}^{factor})^{-1}} = \sum_{i=1}^{n_x} \left(\frac{\hat{x}_i^{true} - \hat{x}_i^{sur}}{\hat{x}_i^{factor}}\right)^2 \tag{35}$$

$$\|\widehat{\boldsymbol{u}}^{true} - \widehat{\boldsymbol{u}}^{sur}\|^2_{\text{diag}(\widehat{\boldsymbol{u}}^{factor})^{-1}} = \sum_{i=1}^{n_u} \left(\frac{\hat{u}_i^{true} - \hat{u}_i^{sur}}{\hat{u}_i^{factor}}\right)^2 \tag{36}$$

where $\hat{x}^{true}$, $\hat{u}^{true}$ are state/control trajectories given by a solution of the mechanistic NMPC problem, $\hat{x}^{sur}$, $\hat{u}^{sur}$ are state/control trajectories given by a solution of the surrogate model NMPC, and $\hat{x}^{factor}$, $\hat{u}^{factor}$ are the normalization factors used to scale the relative magnitude of each state and control variable. Figure 11 shows the value of these metrics for ECE-FS and EFE on Benchmark 1. Again, the solutions obtained by either embedding method are indistinguishable when the same model is used. The PICNN is also able to obtain solutions that are closer to the mechanistic model. Similar results are exhibited for Benchmarks 2 and 3 and are omitted for brevity.

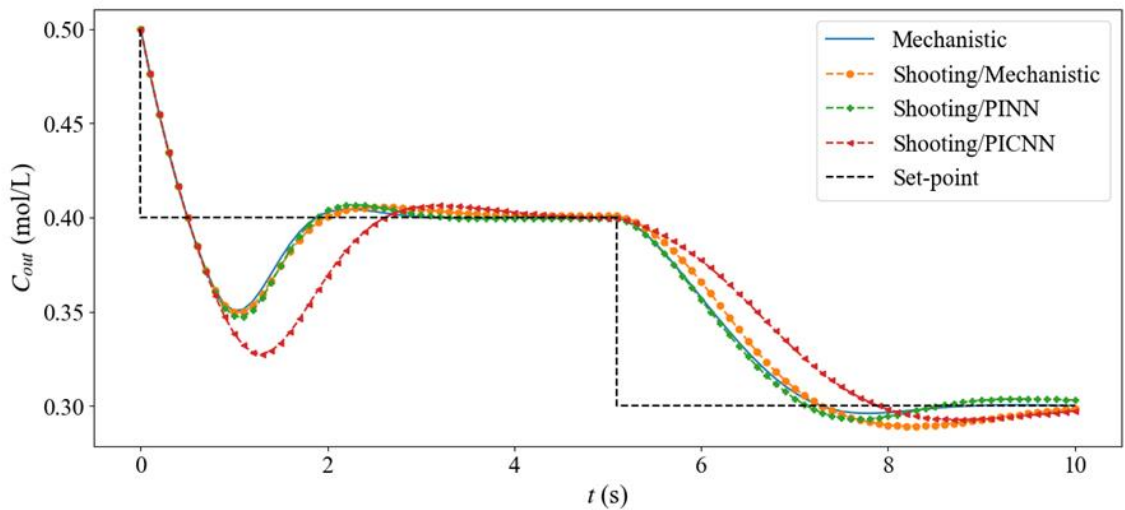

**Figure 12 Setpoint tracking performance of shooting method using different models for Benchmark 1.**

Figure 12 shows the setpoint tracking performance of the shooting method results against the transcribed mechanistic NMPC results. The trajectories exhibit far more variability relative to the direct transcription results. Results obtained with the PICNN deviate from the trajectory to the greatest extent also in contrast to the direct transcription results. These discrepancies can be likely attributed the Jacobians, which are computed via numerical methods instead of AD, leading to different local minima.

## 5. Conclusions

This study benchmarks different NN surrogate embedding strategies in the context of PDE-constrained NMPC problems solved via direct transcription. The results show, for continuous NLPs, that embedding NN surrogates as external functions which evaluate gradients via an ML library (i.e., EFE) can significantly outperform embedding strategies that transform the NN directly into explicit algebraic constraints in the AML where the NMPC problem is modelled (i.e., ECE). Moreover, the results obtained with the case studies considered in this work show that it is not always computationally advantageous to replace mechanistic PDE models with physics-informed NN surrogates, especially for continuous NMPC problems that are solved via direct transcription. Nevertheless, there may be cases where embedding a surrogate within an NMPC may be suitable, e.g., when an accurate mechanistic model is not available. In addition, it has been shown that reformulating nonlinear constraints in mixed-integer NLPs as *ReLU* NNs can have computational advantages [30], [61].

In future work, it would be worthwhile to investigate the performance of ECE and EFE on ODE-constrained NMPC problems solved via direct transcription to examine if a PINN can provide a computational benefit similar to that reported in [15]. Emerging GPU-based NLP solvers, such as that reported in [13], may also help accelerate the performance of surrogate modelling in PDE-constrained NMPC. Finally, alternative NN surrogate architectures should be investigated such as NNs that learn the entire prediction horizon rather than a single sampling step.

## 6. Acknowledgements

The authors gratefully acknowledge the University of Waterloo for its financial support through the Engineering Excellence Master's Fellowship program and to the Natural Sciences & Research Counsil of Canada (NSERC).

## 7. Nomenclature

| Symbol | Description |
|---|---|
| $A$ | reactor cross-sectional area |
| $\boldsymbol{b}_l$ | bias vector at layer $l$ |
| $C$ | concentration, mol L$^{-1}$ |
| $C_{p,s}$ | specific heat capacity of component $s$, J mol$^{-1}$ K$^{-1}$ |
| $\boldsymbol{E}_l$ | dense or linear layer weight matrix at layer $l$ |
| $F_s$ | molar flow rate of component $s$, mol s$^{-1}$ |
| $g$ | path constraints |
| $loss$ | loss function |
| $\boldsymbol{L}$ | matrix of weights on the output variables |
| $M$ | control horizon |
| $NN_\theta$ | neural network parametrized by $\theta$ |
| $P$ | prediction horizon |
| $P_s$ | partial pressure of component $s$, kPa |
| $R$ | ideal gas constant, J mol$^{-1}$ K$^{-1}$ |
| $RR'_j$ | rate of reaction $j$, mol g$^{-1}$ s$^{-1}$ |
| $r_l$ | output of FNN layer $l$ |
| $r_{l,v}$ | output of CNN layer $l$ at spatial location $v$ |
| $\boldsymbol{u}^L, \boldsymbol{u}^U$ | bounds on the manipulated variables |
| $\boldsymbol{u}$ | Manipulated variables |
| $\boldsymbol{W}$ | matrix of weight on the input variables |
| $\boldsymbol{w}$ | EFE model constraints |
| $\boldsymbol{x}^L, \boldsymbol{x}^U$ | bounds on the states |
| $\boldsymbol{x}$ | state variables |
| $\boldsymbol{y}$ | EFE model decision variables |

| Greek Symbol | Description |
|---|---|
| $\mathcal{B}$ | boundary conditions operator |
| $\Delta \boldsymbol{u}_k$ | change in the manipulated variable at time interval $k$ |
| $\mathcal{F}_z$ | nonlinear PDE differential operator |
| $\mathcal{I}$ | initial conditions operator |
| $\nu$ | stoichiometric coefficient |
| $\rho_c$ | catalyst packed density, g cm$^{-3}$ |
| $\sigma$ | activation function |

# A Comparison of Strategies to Embed Physics-Informed Neural Networks in Nonlinear Model Predictive Control Formulations Solved via Direct Transcription

Carlos Andrés Elorza Casas, Luis A. Ricardez-Sandoval, and Joshua L. Pulsipher[1]

*Department of Chemical Engineering, University of Waterloo, Waterloo, Canada, N2L 3G1*

Supplementary Material

Sections:

A. Additional Description of Benchmark Models

B. Training Curves of Benchmark Models

[1] Corresponding author information: Phone: 1-519-888-4567 ext. 42290. Fax: 1-519-888-4347
E-mail: pulsipher@uwaterloo.ca

## A. Additional Description of Benchmark Models

Table A.1 summarizes the reactor specifications and variable bounds for Benchmark 3.

**Table A.1 Variable bounds and reactor specifications for benchmark model 3.**

| Parameter or Variable | Nominal Value | Lower Bound | Upper Bound |
|---|---|---|---|
| $F_{CH_4,in}$ (mmol s$^{-1}$) | 3.11 | 1.56 | 6.22 |
| $F_{H_2,in}$ (mmol s$^{-1}$) | 9.33 | 4.67 | 18.67 |
| $T$ (K) | 848 | 838 | 858 |
| $z$ (cm) | -- | 0.0 | 2.2 |
| $A$ (cm$^2$) | 0.785 | -- | -- |
| $\rho_c$ (g cm$^{-3}$) | 2.835 | -- | -- |
| $P_{tot}$ (atm) | 10.0 | -- | -- |

For Benchmark 3, Equations (A.1)-(A.3) model the kinetics of the reactions occurring in the steam reformer. Here, $RR'_j$ is the rate of reaction $j$ based on catalyst weight. The rate constant of reaction $j$ is given by $k_{rxn,j}$.

$$RR'_1 = \frac{k_{rxn,1}}{P_{H_2}^{2.5}}\left(P_{CH_4}P_{H_2O} - \frac{P_{H_2}^3 P_{CO}}{K_1}\right)/(DEN)^2 \quad (A.1)$$

$$RR'_2 = \frac{k_{rxn,2}}{P_{H_2}}\left(P_{CO}P_{H_2O} - \frac{P_{H_2}P_{CO_2}}{K_2}\right)/(DEN)^2 \quad (A.2)$$

$$RR'_3 = \frac{k_{rxn,3}}{P_{H_2}^{3.5}}\left(P_{CH_4}P_{H_2O}^2 - \frac{P_{H_2}^4 P_{CO_2}}{K_3}\right)/(DEN)^2 \quad (A.3)$$

Equation (A.4) defines the $DEN$ term which depends on the absorption constants $K_{a,s}$. The equilibrium constants of reaction $j$ are given by $K_j$. The Arrhenius equation, (A.5) and (A.6), models the dependence on temperature of the rate constants and adsorption constants, respectively.

$$DEN = 1 + K_{a,CH_4}P_{CH_4} + K_{a,H_2}P_{H_2} + K_{a,CO}P_{CO} + K_{a,H_2O}P_{H_2O}/P_{H_2} \tag{A.4}$$

$$k_{rxn,j} = k^0_{rxn,j} \exp\left(-\frac{E_{A_j}}{RT}\right) \quad \forall j \in \{1,2,3\} \tag{A.5}$$

$$K_{a,s} = B_{a,s} \exp\left(-\frac{\Delta H_{a,s}}{RT}\right) \quad \forall s \in \{CH_4, H_2O, H_2, CO\} \tag{A.6}$$

The Van't Hoff equation, (A.7), models the temperature dependence of the equilibrium constant. The constant $K_{0,j}$ is the equilibrium constant at the reference temperature, $T_{ref}$, which can be calculated from the standard Gibb's free energy of reaction $\Delta G^0_j$ via equation (A.8). The constant $\Delta G^0_j$ can be calculated from the Gibb's free energy of formation, $\Delta G^f_s$, via equation (A.9). The enthalpy of reaction $\Delta H_j$ is calculated from the standard enthalpy of reaction $\Delta H^0_j$ and the heat capacities of each component $C_{p,s}$ via equation (A.10). The constant $\Delta H^0_j$ is calculated from the enthalpies of formation $\Delta H^f_s$ via equation (A.11). The dependence on temperature of the heat capacities is given by equation (A.12).

$$K_j = K_{0,j} \exp\left(\frac{1}{R}\int_{T_{ref}}^{T} \Delta H_j dT\right) \quad \forall j \in \{1,2,3\} \tag{A.7}$$

$$K_{0,j} = \exp\left(-\frac{\Delta G^0_j}{RT_{ref}}\right) \quad \forall j \in \{1,2,3\} \tag{A.8}$$

$$\Delta G^0_j = \sum_s v_{js} \Delta G^f_s \quad \forall j \in \{1,2,3\} \tag{A.9}$$

$$\Delta H_j = \Delta H^0_j + \int_{T_{ref}}^{T} \sum_s v_{js} C_{p,s} \, dT \quad \forall j \in \{1,2,3\} \tag{A.10}$$

$$\Delta H^0_j = \sum_s v_{js} \Delta H^f_s \quad \forall j \in \{1,2,3\} \tag{A.11}$$

$$C_{p,s} = \alpha_s + \beta_s T + \gamma_s T^2 + \delta_s T^3 + \varepsilon_s T^4 \quad \forall s \in \{CH_4, H_2O, H_2, CO_2, CO\} \tag{A.12}$$

To reduce the space of feasible initial conditions for the steam reformer model, the steady-state flow rate profile at the nominal operation is treated as the nominal initial condition. Then, the initial state space of each component is restricted to be between a lower and an upper profile, i.e., $F_{s,0}^L(z) \leq F_{s,0}(z) \leq F_{s,0}^U(z)$, which are determined by varying the control variables between their lower and upper bounds. Figure A.1 shows the bounds on the initial states of the steam reforming model.

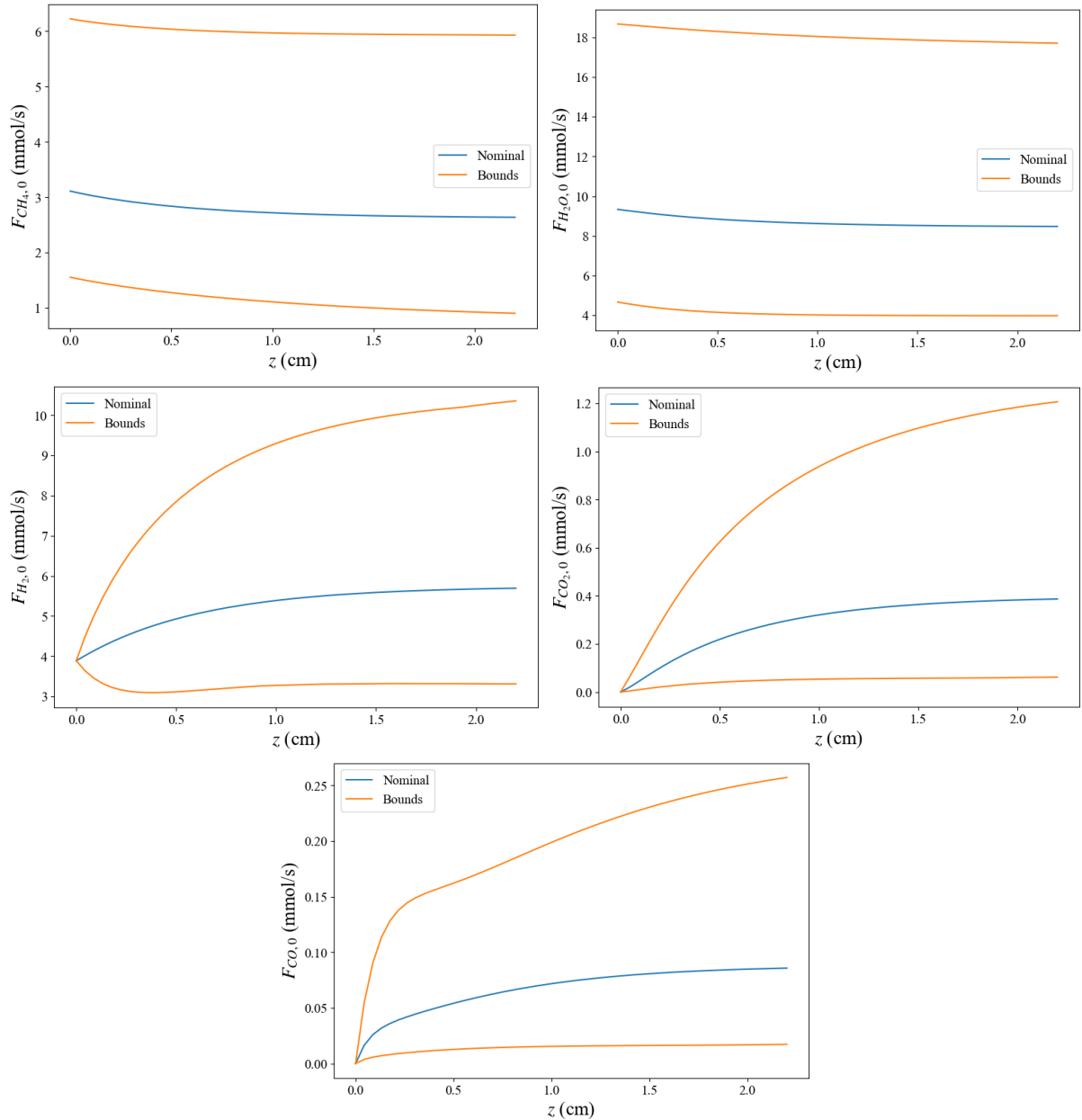

**Figure A.1 Initial state bounds for steam reforming model.**

## B. Additional Description of Benchmark Models

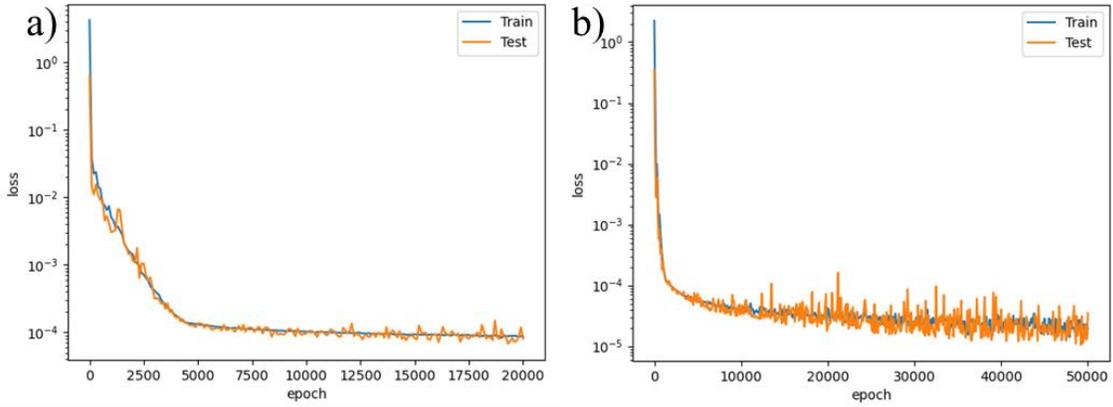

**Figure B.1 PINN, a), and PICNN, b), training curves for Benchmark 1.**

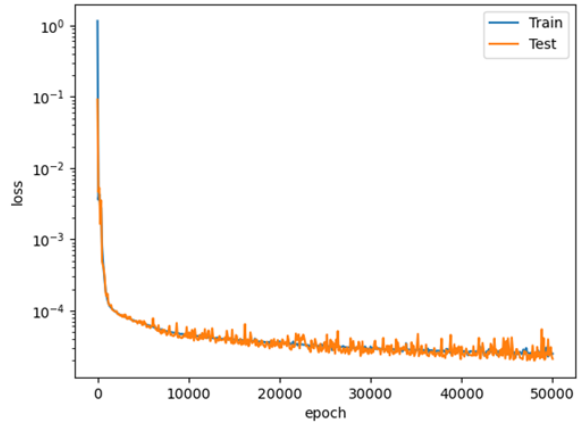

**Figure B.2 PICNN training curve for Benchmark 2.**

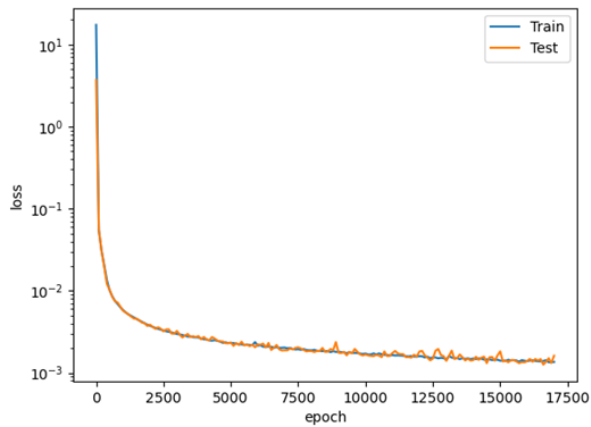

**Figure B.3 PICNN training curve for Benchmark 3.**